\documentclass[reprint,superscriptaddress]{revtex4-1}
\usepackage[utf8]{inputenc}

\usepackage{amsmath}
\usepackage{amssymb}
\usepackage{bm}
\usepackage{graphicx}
\usepackage{xcolor}
\usepackage{hyperref}
\usepackage{ulem}
\usepackage{enumitem}

\newcommand{\p}{\partial}

\newcommand{\e}{\bm{\hat{e}}}

\newcommand{\magn}{\bm{m}}
\newcommand{\St}{S}
\newcommand{\Energy}{E}
\newcommand{\Eex}{E_{\rm ex}}
\newcommand{\Est}{\Energy_{\rm ST}}
\newcommand{\EnergyExt}{F}
\newcommand{\heff}{\bm{f}}

\newcommand{\anisotropy}{\kappa}
\newcommand{\hext}{h}
\newcommand{\bhext}{\bm{h}}
\newcommand{\storque}{\beta}
\newcommand{\pol}{\bm{p}}
\newcommand{\bOmega}{\overline{\Omega}}
\newcommand{\bPsi}{\overline{\Psi}}
\newcommand{\Mminus}{M^-}

\begin{document}

\title{Non-Hermitian dynamics of a two-spin system with $\mathcal{PT}$ symmetry}
\author{Stavros Komineas}
\affiliation{Department of Mathematics and Applied Mathematics, University of Crete, 70013 Heraklion, Crete, Greece}
\affiliation{Institute of Applied and Computational Mathematics, FORTH, 70013 Heraklion, Crete, Greece}
\date{\today}

\begin{abstract}
A system of interacting spins that are under the influence of spin-polarized currents can be described using a complex functional, or a non-Hermitian (NH) Hamiltonian.
We study the dynamics of two exchange-coupled spins on the Bloch sphere.
In the case of currents leading to $\mathcal{PT}$ symmetry, an exceptional point that survives also in the nonlinear system is identified.
The nonlinear system is bistable for small currents and it exhibits stable oscillating motion or it can relax to a fixed point.
The oscillating motion of the two spins is akin to synchronised spin-torque oscillators.
For the full nonlinear system, we derive two conserved quantities that furnish a geometric description of the spin trajectories in phase space and indicate stability of the oscillating motion.
Our analytical results provide tools for the description of the dynamics of NH systems that are defined on the Bloch sphere.
\end{abstract}

\maketitle

\section{Introduction}
\label{sec:introduction}

Effective non-Hermitian (NH) Hamiltonians \cite{2018_NatPhys_ElGanainyMakris,2021_RevModPhys_Bergholtz} have been employed for the description of classical \cite{2016_RMP_KonotopJiankeZezyulin,2019_Science_MiriAlu} and quantum systems \cite{2007_RepProgPhys_Bender,2009_JPA_Rotter} that are coupled to the environment by dissipative or other forces.
The observation that NH Hamiltonians that are invariant under the combination of parity and time reversal ($\mathcal{PT}$) foster real spectra \cite{1998_PRL_BenderBoettcher,2007_RepProgPhys_Bender} has led to a series of theoretical and experimental studies.
Significant amount of work has focused on optical systems with balanced gain and loss such as in optical beam propagation \cite{2010_NatPhys_RueterMakris} and single-mode lasers \cite{2014_Science_FengWongMa}.
More recently, magnetic systems with $\mathcal{PT}$ symmetry have been proposed \cite{2016_PRB_GaldaVinokur,2017_SciRep_GaldaVinokur,2019_SciRep_GaldaVinokur,2020_PhysD_Barashenkov} and systems with gain and loss  \cite{2018_PRL_YangWang,2021_PRAppl_WangGuoBerakdar,2021_arXiv_WittrockPerna,2022_arXiv_DengLiFlebus}, with the latter reports focusing on degeneracies, called exceptional points (EP), in the spectrum of the linear system, were the sensitivity of the system is enhanced.
Effects of EPs were demonstrated for magnonic systems which possess two different loss factors \cite{2019_SciAdv_LiuSun}.


Spin systems are studied extensively in relation to magnetic materials \cite{HubertSchaefer} motivated by applications in magnetic recording, nanoscopic sensors, antennas \cite{2017_NatRevMater_FertReyrenCros}, computing applications and neural network implementations \cite{2016_IEEE_GrollierQuerliozStiles}.
Probing ferromagnets (and other magnetic materials) is done most efficiently by spin torques that are due to a spin-polarized current.
This acts on the magnetic moments in the material in a way that may combine energy gain and loss and it drives magnetization dynamics \cite{2008_JMMM_BerkovMiltat}.
Spin-transfer torque nano-oscillators (STNO) can be constructed that give rise to magnetization oscillators due to injection of dc spin-polarized current \cite{RussekRippardCecil_2010}.
A $\mathcal{PT}$-symmetric system of this kind with a single spin was studied in Ref.~\cite{2016_PRB_GaldaVinokur}. 
The synchronization of chains of STNOs in order to produce large power is a main challenge in this area \cite{2017_NatPhys_AwadAkerman,2017_NatComms_LebrunCros}.

We propose a $\mathcal{PT}$-symmetric system of two interacting spins or magnetic moments that can realistically be constructed when we invoke suitable spin torque effects.
We show that this is described by a complex function instead of a real Hamiltonian.
This gives a paradigm of a realistic nontrivial system on the Bloch sphere.
We show that the nonlinear system exhibits oscillating motion that is connected with the real eigenvalues of the linearized system.
For the full nonlinear system, we derive two conserved quantities that make an explicit connection between the NH system and the Hamiltonian one, and they also furnish a geometric description of the spin trajectories in phase space.

Single spin systems are crucial for quantum computing, and observations on a $\mathcal{PT}$-symmetric single-spin system have been reported \cite{2019_Science_WuLiu}.
An effective spin Hamiltonian is obtained in various physical systems.
An example is a spin Hamiltonian obtained for qubits in an exciton polariton condensate, which are externally controllable by applied laser pulses and are coupled via a coherent tunneling term \cite{2020_NPJ_GhoshLiew}.
Finally, it is well-known that the possible quantum states for a single qubit can be represented on a Bloch sphere.
Our analytical results on complex Hamiltonians could provide tools for a more complete and elegant description of the dynamics of interacting spins on the Bloch sphere in magnetic and other systems.

The paper is organised as follows.
In Sec.~\ref{sec:nonHermitian}, we give the formulation of a spin system using a complex functional.
In Sec.~\ref{sec:twoSpins}, we give the complex function for a system of two spins coupled by exchange.
In Sec.~\ref{sec:nonlinearOscillations}, we give analytically simple periodic solutions for the spin dynamics.
In Sec.~\ref{sec:conservedQuantities}, we derive two conserved quantities and give their geometric meaning.
Sec.~\ref{sec:conclusions} contains our concluding remarks.

\section{A complex function for spins}
\label{sec:nonHermitian}

We consider a spin system described via the magnetization vector $\magn=(m_x, m_y, m_z)$ assumed to have a fixed length normalized to unity, $|\magn|=1$.
If $\Energy$ denotes the magnetic energy, the conservative torque on the magnetization is $\heff=-\p\Energy/\p\magn$.
A polarized spin current that is injected in the system may produce an additional torque and the normalised equation of motion is \cite{2008_JMMM_BerkovMiltat}
\begin{equation}  \label{eq:llgs}
\frac{\p\magn}{\p t} = \magn\times\frac{\p\Energy}{\p\magn} + \alpha\,\magn\times \frac{\p\magn}{\p t}
 -\storque\,\magn\times(\magn\times\pol),
\end{equation}
where $\pol$ is the spin current polarization, $\storque$ is a parameter proportional to the polarized current, and we have included a Gilbert damping term with parameter $\alpha$.
In the case of a ferromagnetic material, the time variable is scaled to $1/(\gamma\mu_0 M_s)$ where $M_s$ is the saturation magnetization and $\gamma$ is the gyromagnetic ratio.

The stereographic projection from the unit sphere $|\magn|=1$ to the complex plane is given by the complex variable
\begin{equation} \label{eq:Omega}
    \Omega = \frac{m_x + i m_y}{1 + m_z}.
\end{equation}
This can be inverted to give the magnetization components
\begin{equation} \label{eq:Omega2m}
    m_x = \frac{\Omega+\bOmega}{1+\Omega\bOmega},\;\; m_y = \frac{1}{i} \frac{\Omega-\bOmega}{1+\Omega\bOmega},\;\; m_z = \frac{1-\Omega\bOmega}{1+\Omega\bOmega},
\end{equation}
where the bar denotes complex conjugation.
The stereographic projection variable will make manifest the non-Hermiticity of the model and it will be central in the formulation of the present work.

The equation of motion for $\Omega$, equivalent to Eq.~\eqref{eq:llgs}, reads
\begin{equation} \label{eq:eqMotion_Omega-0}
(i+\alpha)\,\dot{\Omega} = -\frac{1}{2}\,(1+\Omega\bOmega)^2\,\left( \frac{\p \Energy}{\p \bOmega}
+ i \storque\,\frac{\p \Est}{\p \bOmega} \right)
\end{equation}
where we have defined the function
\begin{equation}
    \Est = - \magn\cdot\pol.
\end{equation}
Formula \eqref{eq:eqMotion_Omega-0} suggests the definition of a complex function that includes the spin torque term,
\begin{equation} \label{eq:complexEnergy}
    \EnergyExt = \Energy + i\storque\Est,
\end{equation}
so that the equation of motion for $\Omega$ takes the compact form
\begin{equation} \label{eq:eqMotion_Omega-1}
(i+\alpha)\,\dot{\Omega} = -\frac{1}{2}\,(1+\Omega\bOmega)^2\, \frac{\p \EnergyExt}{\p \bOmega}.
\end{equation}
The function $F$ will be called the complex Hamiltonian.

As a basic example, if we have an external magnetic field $\bhext=(0,0,\hext)$ giving rise to the energy $\Energy=-\magn\cdot\bhext$, and the spin polarization is $\pol=(0,0,1)$, then $F=-(\hext + i \storque) m_z$ and the equation of motion is
\begin{equation}
    (i+\alpha)\,\dot{\Omega} = -(\hext + i\storque) \Omega.
\end{equation}

\section{Two exchange-coupled spins}
\label{sec:twoSpins}

We proceed to consider two magnetization vectors or spins interacting via exchange.
These may represent two domains or layers in a magnetic material \cite{2020_NatMater_LegrandCrosFert}, or two effective spins in another physical system. 

We denote the two vectors by $\magn_1, \magn_2$ and assume that they have equal length, $|\magn_1|=|\magn_2|=1$.
Except for the exchange interaction, we include an easy-axis anisotropy along $z$, and an external field  $\bhext=\hext\e_z$.
The energy is
\begin{equation} \label{eq:energy}
   \Energy = -J\, \magn_1\cdot\magn_2 - \frac{\anisotropy}{2} \left[ \left( m_{1,z} \right)^2 + \left( m_{2,z} \right)^2 \right] - \hext\,
   (m_{1,z}+m_{2,z})
\end{equation}
where $J > 0,\,\anisotropy > 0$ are the exchange and anisotropy parameters respectively, and $m_{1,z}, m_{2,z}$ denote the $z$ components of the two vectors.
We further consider that spin-polarized currents are injected in the two domains $\magn_1, \magn_2$, with polarization in two opposite directions, $\pol = \pm \e_z$.
(Such a strategy for achieving $\mathcal{PT}$ symmetry is mentioned in \cite{2016_RMP_KonotopJiankeZezyulin} and attributted to \cite{2016_Gaididei}.)
The opposite polarizations could be achieved, e.g., in the case of two coupled ferromagnetic layers, by injecting a current through the layer separating the two magnetic layers \cite{2021_PRAppl_WangGuoBerakdar}, or by two different currents through layers adjacent to each one of the magnetic layers.
Spin current polarization with a component perpendicular to the sample plane was recently demonstrated \cite{2022_NatElectr_BoseSchreiberRalph,2022_NatElectr_BoseRalph}.
The equations of motion are
\begin{equation} \label{eq:twoSpins_dynamics}
\begin{split}
\dot{\magn}_1 & = -\magn_1\times \bm{f}_1 + \alpha \magn_1\times\dot{\magn}_1 - \storque \magn_1 \times (\magn_1\times \e_z)  \\
\dot{\magn}_2 & = -\magn_2\times \bm{f}_2 + \alpha \magn_2\times\dot{\magn}_2 + \storque \magn_2 \times (\magn_2\times \e_z)
\end{split}
\end{equation}
where the dot denotes time differentiation and the effective fields are
\begin{equation} \label{eq:effectiveFields}
\begin{split}
\bm{f}_1 & = -\frac{\p \Energy}{\p \magn_1} = J\magn_2 + \anisotropy\, m_{1,z} \e_z + \hext\,\e_z, \\
\bm{f}_2 & = -\frac{\p \Energy}{\p \magn_2} = J\magn_1 + \anisotropy\, m_{2,z} \e_z + \hext\,\e_z.
\end{split} \notag
\end{equation}

We will assume in the following $\hext>0,\, \storque > 0$.
Eqs.~\eqref{eq:twoSpins_dynamics} have four fixed points.
Two fixed points correspond to both $\magn_1, \magn_2$ pointing at the north or at the south pole,
\begin{align} \label{eq:P1}
    & \magn_1=\magn_2=\e_z \tag{P1} \\
\label{eq:P2}
    & \magn_1=\magn_2 = -\e_z \tag{P2}
\end{align}
and two further fixed points correspond to the vectors pointing at opposite poles,
\begin{align}
\label{eq:P3}
    & \magn_1=\e_z,\quad\;\;\, \magn_2=-\e_z \tag{P3} \\
    \label{eq:P4}
    & \magn_1=-\e_z,\quad \magn_2=\e_z. \tag{P4}
\end{align}

For the study of the stability of the fixed points as well as for using the concepts of non-Hermiticity, we proceed to the formulation of the system using the stereographic projections \eqref{eq:Omega} of each of the spins, denoted by $\Omega_1, \Omega_2$.
We write the complex Hamiltonian of Eq.~\eqref{eq:complexEnergy} for the system of two spins,
\begin{equation} \label{eq:energyExt}
\begin{split}
\EnergyExt = & -J\, \magn_1\cdot\magn_2 - \frac{\anisotropy}{2} \left[ \left( m_{1,z} \right)^2 + \left( m_{2,z} \right)^2 \right] \\
 & - \hext\,(m_{1,z}+m_{2,z}) + i\storque\,(m_{2,z}-m_{1,z}).
\end{split}
\end{equation}
This can be written in terms of the stereographic variables using Eqs.~\eqref{eq:Omega2m}.
For example, the exchange term is
\begin{equation} \label{eq:magn1magn2}
    -J\magn_1\cdot\magn_2 = 2J \frac{(\Omega_1-\Omega_2)(\bOmega_1-\bOmega_2)}{(1+\Omega_1\bOmega_1)(1+\Omega_2\bOmega_2)}.
\end{equation}
The equations of motion for $\Omega_1, \Omega_2$, from Eq.~\eqref{eq:eqMotion_Omega-1}, read
\begin{equation} \label{eq:twoSpins_Omega}
\begin{split}
(i+\alpha)\, \dot{\Omega}_1 = J \frac{1+\Omega_1\bOmega_2}{1+\Omega_2\bOmega_2} (\Omega_2-\Omega_1)
& -\anisotropy \frac{1-\Omega_1\bOmega_1}{1+\Omega_1\bOmega_1}\,\Omega_1 \\
 & - (\hext+i\storque)\Omega_1  \\
(i+\alpha)\, \dot{\Omega}_2 = J \frac{1+\bOmega_1\Omega_2}{1+\Omega_1\bOmega_1} (\Omega_1-\Omega_2)
 & -\anisotropy \frac{1-\Omega_2\bOmega_2}{1+\Omega_2\bOmega_2}\,\Omega_2 \\
& - (\hext-i\storque)\Omega_2.
\end{split}
\end{equation}
System \eqref{eq:twoSpins_Omega} is $\mathcal{PT}$-symmetric when we neglect the damping term by setting $\alpha=0$.
The terms with $\storque$ act as gain and loss, but note that not each separate one of them can be classified as giving only gain or only loss.
This is unlike in previous works on magnetics \cite{2015_PRB_LeeKottosShapiro,2018_PRL_YangWang,2021_PRAppl_WangGuoBerakdar,2022_arXiv_DengLiFlebus}, where damping and anti-damping torques were assumed \cite{2022_arXiv_HurstFlebus}.

The solution of Eq.~\eqref{eq:twoSpins_Omega} $\Omega_1=0,\, \Omega_2=0$ corresponds to the fixed point \eqref{eq:P1} where both vectors point at the north pole.
We can study the behavior of the system close to the fixed point if we assume $|\Omega_1|, |\Omega_2| \ll 1$ and linearize the equations.
We have the linearized system (for $\alpha=0$)
\begin{equation} \label{eq:linearized_Omega}
\begin{split}
i \dot{\Omega}_1 & = -(\omega_0 + i\storque) \Omega_1 + J\Omega_2,\qquad \omega_0=J+\anisotropy + \hext \\
i \dot{\Omega}_2 & = -(\omega_0 - i\storque) \Omega_2 + J\Omega_1
\end{split}
\end{equation}
that has the form of a standard linear $\mathcal{PT}$-symmetric dimer model.
For $\storque \leq J$, we define an angle $\theta$ via
\begin{equation}  \label{eq:conditionTheta}
\sin\theta = \frac{\storque}{J}, \qquad 0 \leq \theta \leq \pi
\end{equation}
and have a solution
\begin{equation} \label{eq:eigenstates}
  \Omega_1 = A\,e^{i\omega t},\quad \Omega_2 = A\, e^{i\theta}\,e^{i\omega t},  
\end{equation}
with angular frequency
\begin{equation} \label{eq:freq_P1_stable}
\omega = \omega_0 - J\cos\theta.
\end{equation}
This solution corresponds to a periodic motion of both spins with equal amplitudes $|\Omega_1|=|\Omega_2|$.
Equivalently, one can say that the spins have equal $z$ components, $m_{1,z}=m_{2,z}$ while precession occurs around the $z$ axis.

For $\storque > J$, the angular frequency has an imaginary part,
\begin{equation} \label{eq:freq_P1_unstable}
\omega = \omega_0 \pm iJ\sqrt{(\storque/J)^2 - 1},
\end{equation}
which means that $|\Omega_1|, |\Omega_2|$ will generically grow in time and, thus, the fixed point is unstable.
A corresponding analysis for the fixed point \eqref{eq:P2} gives the same stability results, that is, precessional motion for $\storque \leq J$ and an instability for $\storque > J$.

In order to study the fixed point \eqref{eq:P3}, we use the transformation $\Psi_2=1/\bOmega_2$ for the second spin.
Then, \eqref{eq:P3} corresponds to $\Omega_1=0, \Psi_2=0$.
The equations of motion \eqref{eq:twoSpins_Omega} become (for $\alpha=0$)
\begin{equation}  \label{eq:twoMacrospin_OmegaPsi}
\begin{split}
i\, \dot{\Omega}_1 = J \frac{1-\Omega_1\bPsi_2}{1+\Psi_2\bPsi_2} (\Omega_1+\Psi_2)
 & -\anisotropy \frac{1-\Omega_1\bOmega_1}{1+\Omega_1\bOmega_1}\,\Omega_1 - (\hext+i\storque)\Omega_1  \\
i\, \dot{\Psi}_2 = J \frac{\bOmega_1\Psi_2-1}{1+\Omega_1\bOmega_1} (\Omega_1+\Psi_2)
 & -\anisotropy \frac{\Psi_2\bPsi_2-1}{\Psi_2\bPsi_2+1}\,\Psi_2  - (\hext+i\storque)\Psi_2.
\end{split}
\end{equation}
In order to study stability, we assume $\Omega_1, \Psi_2 \ll 1$ and linearize the equations to obtain
\begin{equation}  \label{eq:twoMacrospin_OmegaPsi_linear}
\begin{split}
i\dot{\Omega}_1 & = (J - \anisotropy - \hext - i\storque) \Omega_1 + J \Psi_2 \\
i\dot{\Psi}_2 & = (-J + \anisotropy - \hext - i\storque) \Psi_2 - J \Omega_1.
\end{split}
\end{equation}
Assuming solutions
\[
\Omega_1 = A_1 e^{i\omega t},\quad \Psi_2 = \frac{1}{\bOmega_2} = A_2 e^{i\omega t},
\]
we obtain the condition
\begin{equation}
(\omega - \hext - i\storque)^2 = \anisotropy (\anisotropy - 2J).
\end{equation}
In the case
\begin{equation}
 \anisotropy \geq 2J \Rightarrow \omega = \hext \pm \sqrt{\anisotropy (\anisotropy - 2J)} + i\storque,
\end{equation}
we have that $i\omega$ has a negative real part, $-\storque$, giving asymptotic stability for every $\storque > 0$.
In the case
\begin{equation} \label{eq:P3eigenvalue}
\anisotropy < 2J \Rightarrow \omega = \hext + i \left[ \storque \pm \sqrt{\anisotropy (2J - \anisotropy)} \right],
\end{equation}
we have that
\begin{equation} \label{eq:oppositePoles_stability}
\storque \geq \sqrt{\anisotropy (2J - \anisotropy)}
\end{equation}
gives asymptotic stability, while
\begin{equation} \label{eq:oppositePoles_instability}
\storque <  \sqrt{\anisotropy (2J - \anisotropy)}
\end{equation}
gives instability.
The right side in Eq.~\eqref{eq:P3eigenvalue} has a maximum equal to $J$ at $\anisotropy=J$.
This means that, for $\storque > J$, the fixed point is stable for every value of $\anisotropy$.

Finally, for the fixed point \eqref{eq:P4}, we can follow a similar procedure and find that it is unstable for every value of $\storque > 0,\, \anisotropy > 0$.
Table~\ref{tab:stability} summarizes the results for the stability of the fixed points.

\begin{table}[ht]
    \centering
    \begin{tabular}{|l|c|c|}
            \hline
            & stable & unstable \\
            \hline
         P1 & $\storque \leq J$ & $\storque > J$ \\
         P2 & $\storque \leq J$ & $\storque > J$ \\
         P3 & $\anisotropy \geq 2J$ or $\storque \geq \sqrt{\anisotropy (2J - \anisotropy)}$ & $\anisotropy < 2J$ and $\storque < \sqrt{\anisotropy (2J - \anisotropy)}$ \\
         P4 & - & $\storque > 0$ \\
         \hline
    \end{tabular}
    \caption{Stability regimes for the four fixed points.
    The results of linear stability analysis agree with numerical simulation results for the nonlinear system.
    We are confined to the case $J,\storque > 0$.}
    \label{tab:stability}
\end{table}

The linear stability results of this section determine cases where fixed points are unstable.
On the other hand, in the cases where the linear system is stable (with a real frequency), no conclusive result can be drawn for the nonlinear system (as dictated by standard dynamical systems theory).
Further study of the stability will be given in Sec.~\ref{sec:numerics}.

\section{Nonlinear oscillations}
\label{sec:nonlinearOscillations}

\subsection{Amplitude and frequency}

We proceed to study solutions of the nonlinear system \eqref{eq:twoSpins_Omega}.
We start by assuming solutions of the form \eqref{eq:eigenstates} that is, perfect oscillations where the two spins differ by a phase.
Substituting in Eqs.~\eqref{eq:twoSpins_Omega}, we obtain
\begin{align*}
 |A|^2 & \left[ J (1 - e^{-i\theta}) + \anisotropy + \omega - \hext - i\storque \right] \\
 & + \left[ J (e^{i\theta}-1) - \anisotropy + \omega - \hext - i\storque \right] = 0,  \\
 |A|^2 & \left[ J (1 - e^{i\theta}) + \anisotropy + \omega - \hext + i\storque \right] \\
 & + \left[ J (e^{-i\theta}-1) - \anisotropy + \omega - \hext + i\storque \right] = 0.
\end{align*}
The two equations are identical if condition \eqref{eq:conditionTheta} holds, and they reduce to
\begin{align*}
|A|^2 & \left[ J (1-\cos\theta) + \anisotropy + (\omega - \hext) \right] \\
 + & \left[ J (1-\cos\theta) + \anisotropy - (\omega - \hext) \right] = 0.
\end{align*}
This obtains the angular frequency as a function of the amplitude (as expected for nonlinear oscillators),
\begin{equation} \label{eq:freq_nonlinear}
\omega = \frac{1-|A|^2}{1+|A|^2} \left[ J\, (1-\cos\theta) + \anisotropy \right] + \hext.
\end{equation}
For the interpretation of this result, one should note that $(1-|A|^2)/(1+|A|^2)$ gives the $z$ component of $\magn_1,\magn_2$.
For every $\storque \leq J$, Eq.~\eqref{eq:conditionTheta} gives two angles $\theta_1, \theta_2$ with $\theta_2=\pi-\theta_1$ (assuming $\theta_1 < \pi/2$)
and thus two frequency values are given by \eqref{eq:freq_nonlinear}.
For $0 \leq \theta=\theta_1 \leq \pi/2$, we obtain an acoustic branch and for $\pi/2 \leq \theta=\theta_2 \leq \pi$, we obtain an optical branch.
Related observations were reported in \cite{2019_SciAdv_LiuSun} for a passive magnonic system.

Eq.~\eqref{eq:conditionTheta} implies an exceptional point at $\storque=J$ for the nonlinear system.
This means that the system possesses periodic orbits for $\storque < J$ but these are not sustained for $\storque>J$.
Furthermore, at $\storque=J$, Eq.~\eqref{eq:conditionTheta} gives a single solution $\theta=\pi/2$, the two frequencies coalesce to a single one and we have only one nonlinear oscillation solution \eqref{eq:eigenstates}.
The exceptional point of the nonlinear system coincides with that obtained for the linearized system \eqref{eq:linearized_Omega}.
The stability of the periodic orbits in the nonlinear system will be studied numerically in the next subsection.

\begin{figure*}[t]
    \centering
    (a)\includegraphics[width=5.5cm]{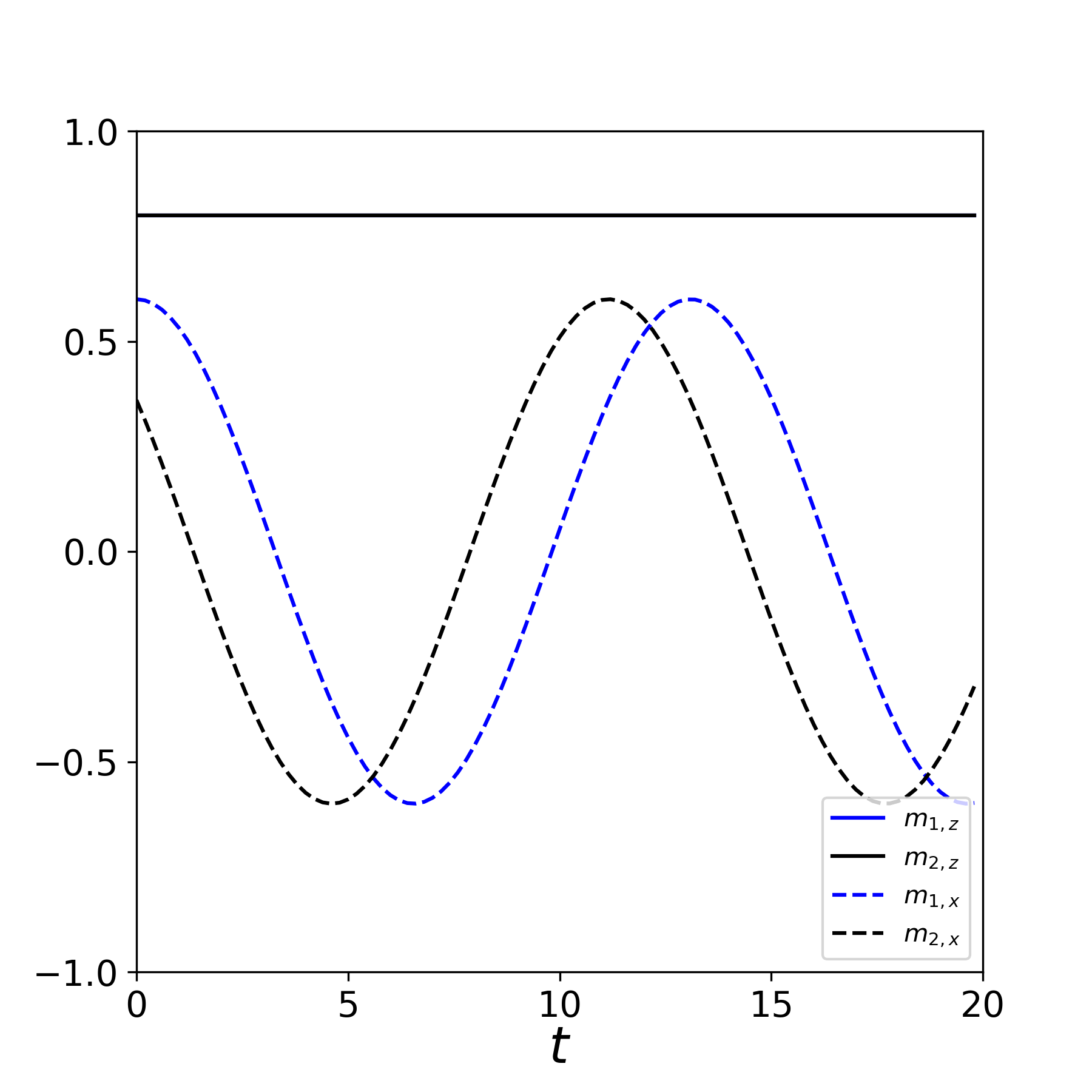}\hspace{1cm}
    (b)\includegraphics[width=5.5cm]{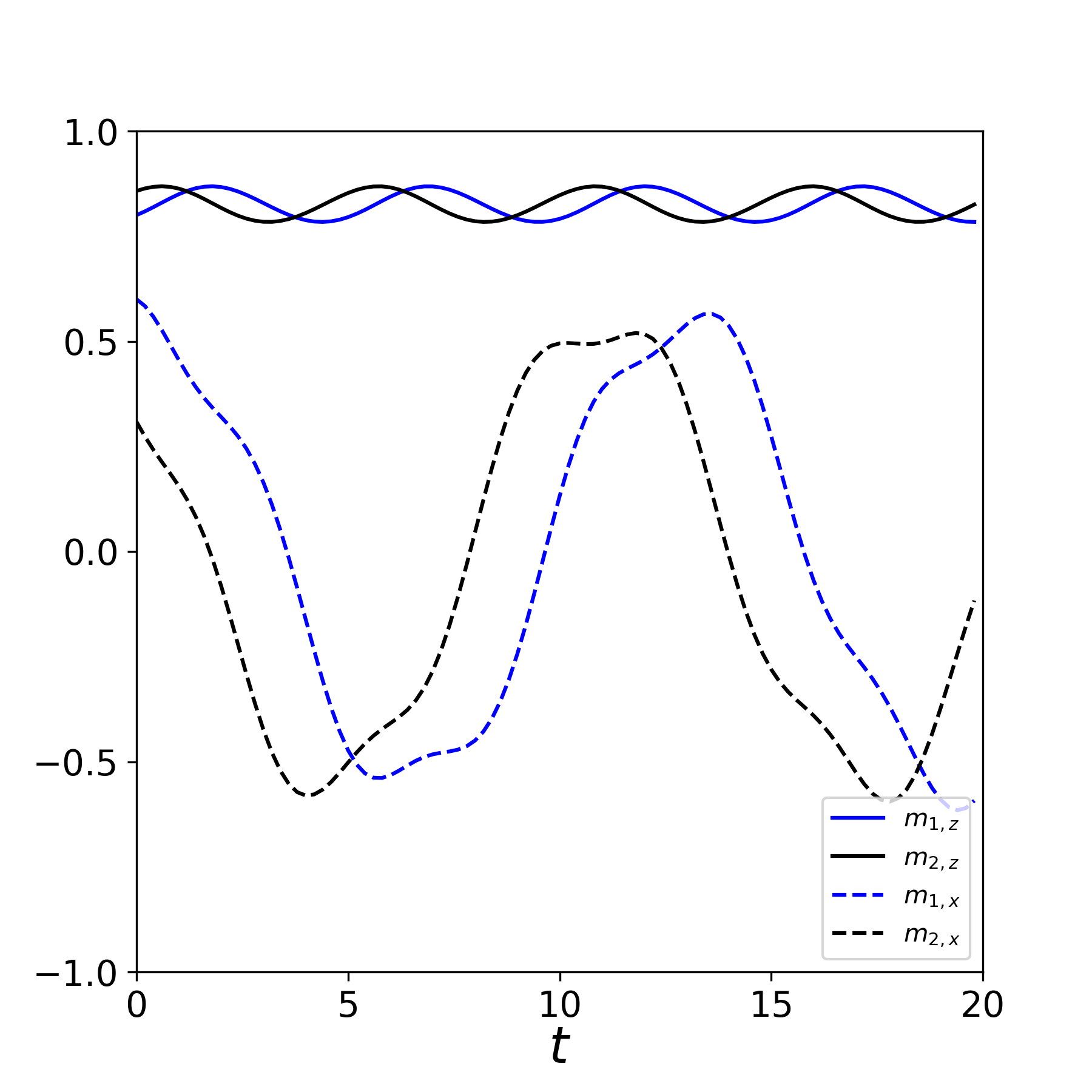}
    (c)\includegraphics[width=5.5cm]{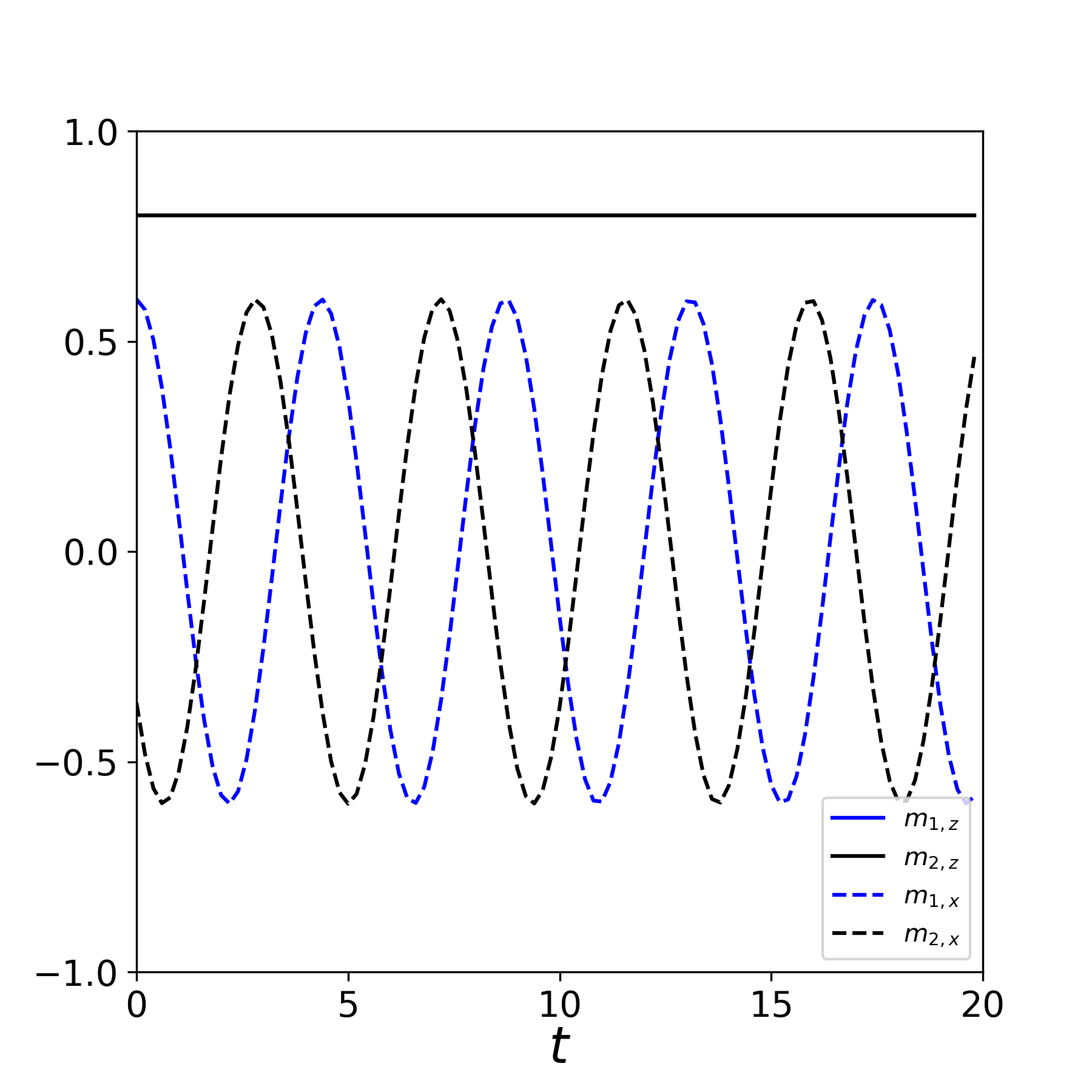}\hspace{1cm}
    (d)\includegraphics[width=5.5cm]{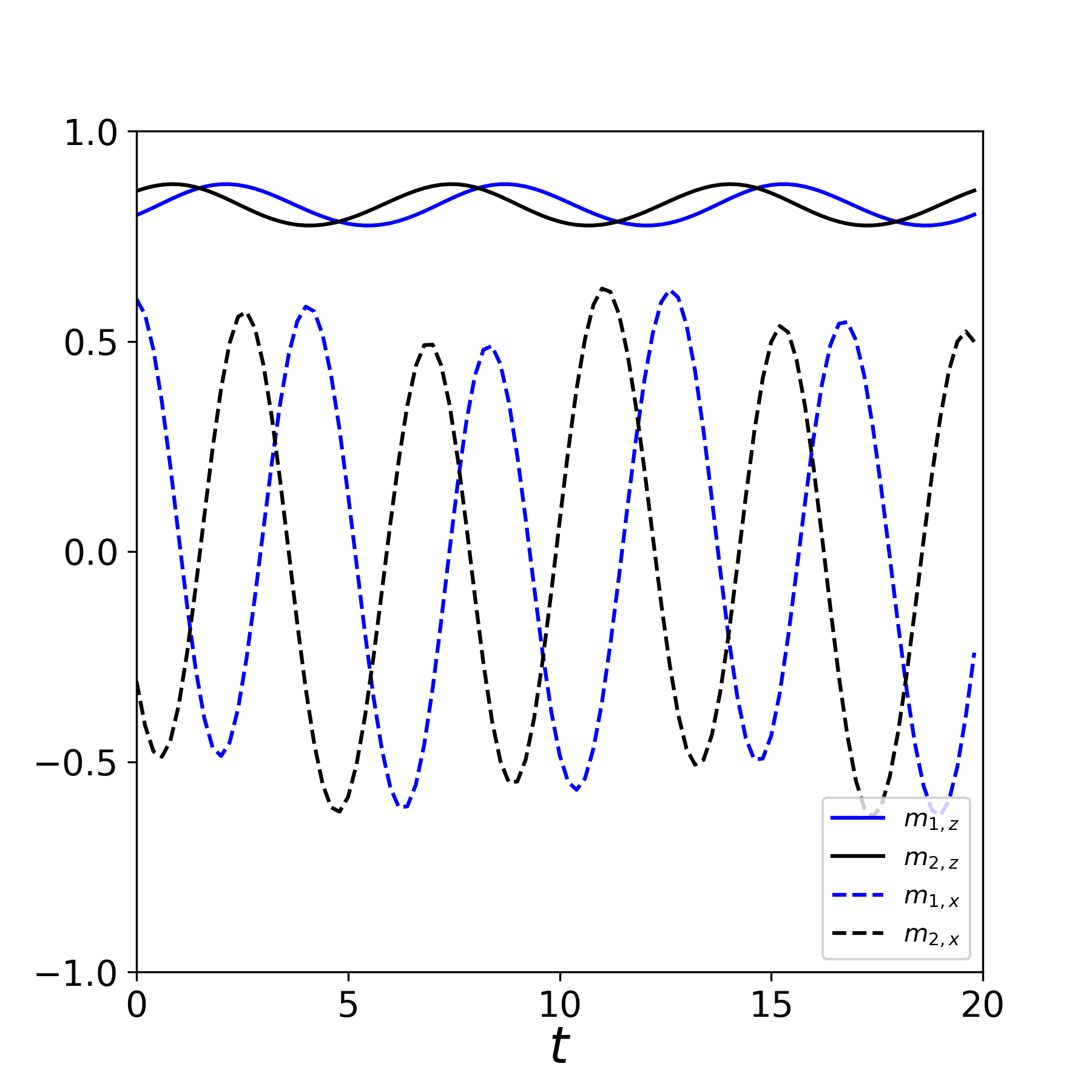}
    \caption{We simulate Eqs.~\eqref{eq:twoSpins_Omega} with parameters $J=1, \anisotropy=0.2, \hext=0$ and spin-torque parameter $\storque=0.8$.
    The first and the third components of the magnetization $m_{1,x}(t), m_{2,x}(t)$ and $m_{1,z}(t), m_{2,z}(t)$ are shown.
    (a) Using an initial condition that agrees with Eqs.~\eqref{eq:conditionTheta} (for $\theta < \pi/2$) and \eqref{eq:eigenstates}, we obtain spin precession in the acoustic branch, that is, oscillations around the nonlinear eigenstate \eqref{eq:eigenstates}.
    (b) When we choose the initial condition close to the values used in (a), we obtain oscillations of the $z$ component of the spins in addition to precessional motion.
    (c) Using an initial condition that agrees with Eqs.~\eqref{eq:conditionTheta} (for $\theta > \pi/2$) and \eqref{eq:eigenstates}, we obtain spin precession in the optical branch.
    (d) Similar simulation to (b) for the optical branch.}
    \label{fig:oscillations}
\end{figure*}
    
\begin{figure}[t]
    \centering
    \includegraphics[width=5.5cm]{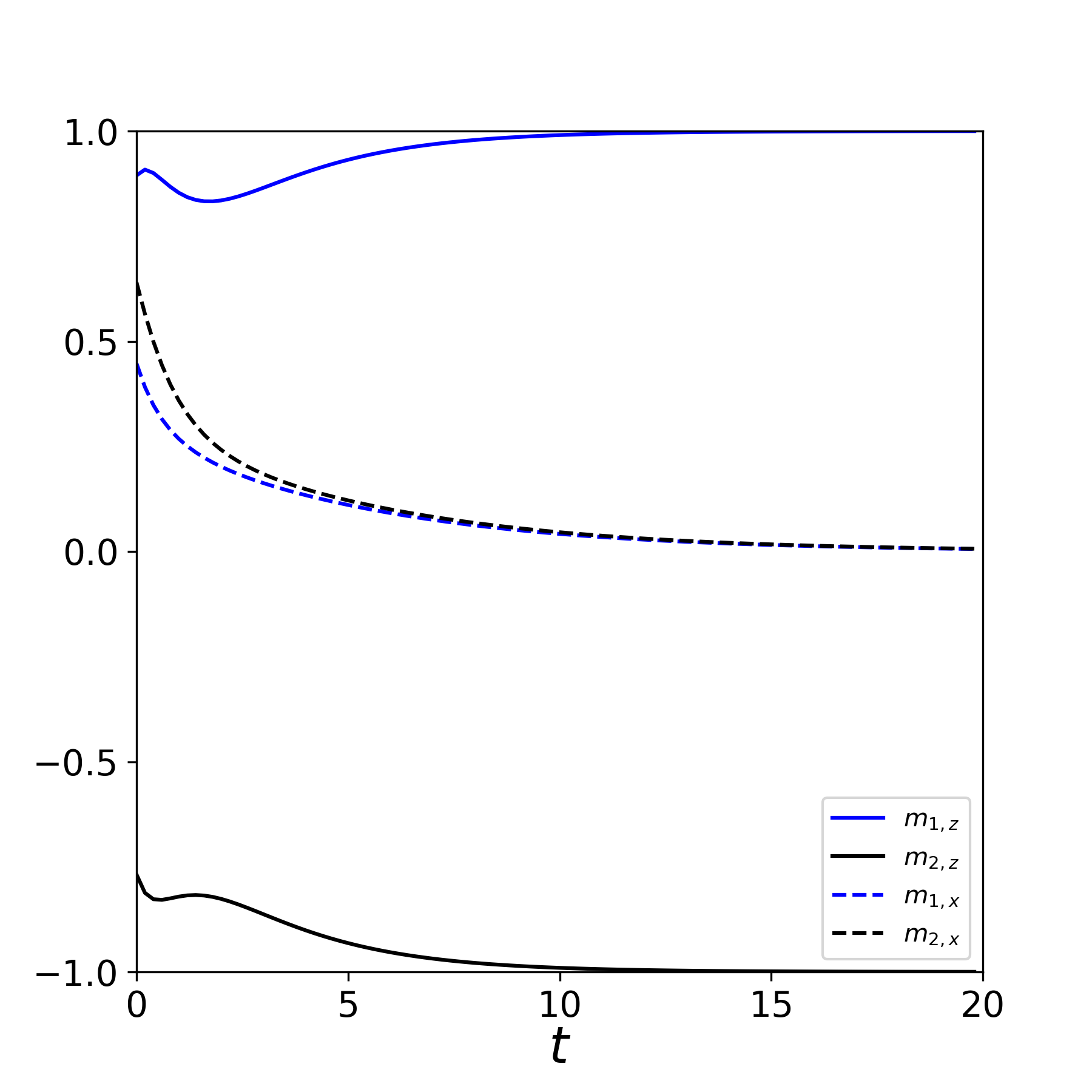}
    \caption{
    We use the same parameter values as in the simulation in Fig.~\ref{fig:oscillations} so that $\storque > \sqrt{\anisotropy (2J-\anisotropy)} = 0.6$.
    We use an initial condition close to the fixed point \eqref{eq:P3}.
    We observe that the system goes asymptotically to this fixed point.
    }
    \label{fig:oppositePoles}
\end{figure}

We conclude the subsection by making some further remarks on Eq.~\eqref{eq:freq_nonlinear} and assuming $\hext=0$ for simplicity.
The frequency is invariant under the transformation $A\to 1/A$.
This corresponds to the invariance of the model under the transformation $m_3\to - m_3$.
For small amplitude, $|A|\to 0$, the spins are close to the north pole of the Bloch sphere and Eq.~\eqref{eq:freq_nonlinear} reduces to the result \eqref{eq:freq_P1_stable} of the linear model.
The limit $|A|\to \infty$ is allowed and it corresponds to the spins being close to the south pole of the sphere.
The frequency of oscillation has maximum absolute value for $\magn$ close to the north ($A\to 0$) or the south pole ($|A|\to\infty$).

In the case that both spins point on the equator, $|A|=1$, there is no spin precession, $\omega=0$, according to Eq.~\eqref{eq:freq_nonlinear} for $\hext=0$.
Therefore, the configurations $\Omega_2 = e^{i\theta}\Omega_1$, for $|\Omega_1|=|\Omega_2| = 1$, give a continuum of fixed points of the system.
These are additional to the four fixed points P1-P4 discussed in Sec.~\ref{sec:twoSpins}.
The new fixed points are obviously unstable, as any deviation from the equator would result in spin precession around the $z$ axis, that means, the spins would go away from their fixed point positions.

\subsection{Numerics}
\label{sec:numerics}

We simulate numerically the system when it is below the exceptional point, for $\storque < J$, and find that it follows the precessional eigenstates \eqref{eq:eigenstates} of the nonlinear system if the initial spin configuration is prepared so that it agrees with Eqs.~\eqref{eq:conditionTheta} and \eqref{eq:eigenstates}.
Fig.~\ref{fig:oscillations}a shows an example of spin precession for the acoustic branch.
Fig.~\ref{fig:oscillations}c shows an example of spin precession for the optical branch.
The graphs show the $x$ and the $z$ components of the magnetization vector.
The frequency of precession is given by Eq.~\eqref{eq:freq_nonlinear} and the result is verified in the graphs by the periodicity of the $x$ components of the magnetization vectors.

Furthermore, for an initial condition that is close to the eigenstate \eqref{eq:eigenstates}, we obtain quasi-periodic motion where the $z$ components of the spins oscillate while the spins precess around the $z$ axis.
An example is shown in Fig.~\ref{fig:oscillations}b, for an initial condition close to the one that gave the acoustic branch periodic motion in Fig.~\ref{fig:oscillations}a.
Another example is shown in Fig.~\ref{fig:oscillations}d, for an initial condition close to the one that gave the optical branch periodic motion in Fig.~\ref{fig:oscillations}c.
Two frequencies are involved in this motion.
The precessional motion frequency is close to \eqref{eq:freq_nonlinear}, while the periodicity of the oscillation of the $z$ component of the spin gives a second frequency apparently unrelated to the spin precession frequency.

For the simulations in Fig.~\ref{fig:oscillations}, we have chosen parameter such that condition \eqref{eq:oppositePoles_stability} is satisfied, and the system is expected to be bistable as inferred from Table~\ref{tab:stability}.
We run a further simulation using the same parameter set, but now choosing an initial condition close to the point \eqref{eq:P3}.
We find that the system goes asymptotically to this fixed point, as shown in Fig.~\ref{fig:oppositePoles}.
This verifies the bistability of the system.

In the case that $\storque$ satisfies condition \eqref{eq:oppositePoles_instability}, only \eqref{eq:P1}, \eqref{eq:P2} are stable.
Starting from any initial condition the system goes into a periodic or a quasi periodic motion similar to those shown in Fig.~\ref{fig:oscillations}.

Finally, when we cross the exceptional point, for $\storque > J$, the fixed points \eqref{eq:P1}, \eqref{eq:P2} are unstable due to the imaginary part in the eigenvalues \eqref{eq:freq_P1_unstable}.
Then, \eqref{eq:P3} is the only stable fixed point and the system goes to that asymptotically from any initial condition.

We conclude the section with a note about the remarkable periodicity in the dynamics of this system.
The quasi-periodic motion observed in many of the simulations, cannot be considered as direct consequencies of the results obtained in Sec.~\ref{sec:twoSpins} for the linearized equations.
They are rather due to a partial or complete integrability of the system.
The periodic and quasi-periodic dynamics can be anticipated due to the existence of the integrals that will be given in Sec.~\ref{sec:conservedQuantities}.

\section{Conserved quantities}
\label{sec:conservedQuantities}

\subsection{Energy and magnetization}
\label{sec:EnergyMagnetization}

The existence of integrals of motion is implied by the correspondence between the $\mathcal{PT}$-symmetric Hamiltonian with a Hermitian one \cite{2003_Mostafazadeh}.
This is further supported by the existence of periodic and quasiperiodic solutions of the system of equations \eqref{eq:twoSpins_dynamics}.
It has been shown that a class of $\mathcal{PT}$-symmetric nonlinear Schr\"odinger dimers admit a Hamiltonian and are completely integrable systems \cite{2014_PRA_Barashenkov,2014_JPA_BarashenkovGianfreda,2015_JPA_BarashenkovPelinovsky}.

We will derive integrals of motion for the undamped ($\alpha=0$) system.
We start by noting that, in the absence of spin torque ($\storque=0$), the energy \eqref{eq:energy} and also the total magnetization along the symmetry axis
\begin{equation} \label{eq:Mz}
    M_z = m_{1,z} + m_{2,z}
\end{equation}
are conserved quantities.
When $\storque \neq 0$, the time derivative of the energy is
\begin{equation} \label{eq:energy_timeDerivative}
\frac{d\Energy}{dt} = -\storque \left( \Energy - J - \frac{\anisotropy}{2} M_z^2 + \anisotropy \right) \Mminus
\end{equation}
and the time derivative of $M_z$ is
\begin{equation}  \label{eq:Mz_timeDerivative}
\frac{d M_z}{dt} = - \storque\, M_z \Mminus,
\end{equation}
where we have defined the quantity
\begin{equation} \label{eq:M-}
    \Mminus = m_{1,z}-m_{2,z}
\end{equation}
that will enter many calculations in this section.

Eqs.~\eqref{eq:energy_timeDerivative}, \eqref{eq:Mz_timeDerivative} suggest that we define the quantity
\begin{equation} \label{eq:G}
G = E - J + \frac{\anisotropy}{2} (M_z^2 + 2)
\end{equation}
whose time derivative has the form
\begin{equation} \label{eq:G_timeDerivative}
\frac{dG}{dt} = - \storque\, G \Mminus.
\end{equation}
Eq.~\eqref{eq:G_timeDerivative}
together with Eq.~\eqref{eq:Mz_timeDerivative} give the conserved quantity
\begin{equation} \label{eq:conserved1}
I_1 = \frac{E - J + \frac{\anisotropy}{2} (M_z^2 + 2)}{M_z}.
\end{equation}

We can make further progress if we now confine ourselves to the {\it exchange model}, i.e., $\anisotropy=0$.
We have the conserved quantity
\begin{equation} \label{eq:conserved1_exchange}
I_1 = \frac{\Eex-J}{M_z}.
\end{equation}
This is valid also in the presence of a field, $\hext\neq 0$.

\subsection{Exchange model, first integral}
\label{sec:firstIntegral}

\begin{figure}[t]
\begin{center}
\includegraphics[width=0.7\columnwidth]{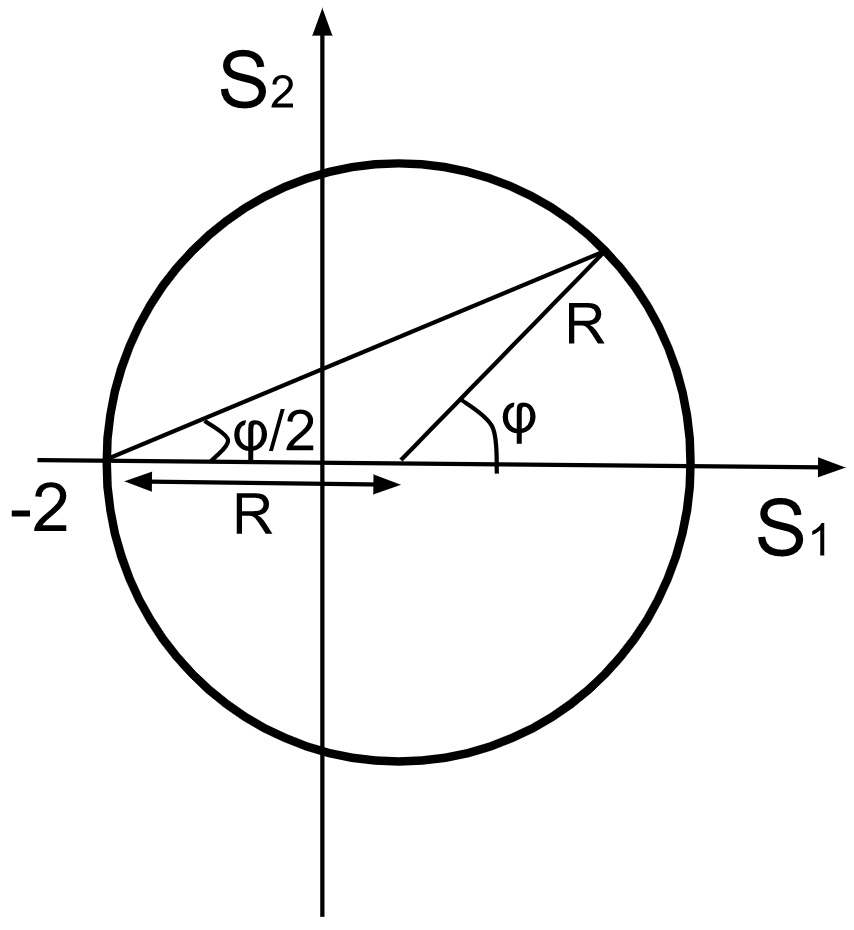}
\caption{The geometry of Eq.~\eqref{eq:S1S2_circle} and Eq.~\eqref{eq:S1S2_circle_parametric}.
Note that $R$ may also be negative (then the center of the circle would be at $S_1 < -2$).}
\label{fig:S1S2_circle}
\end{center}
\end{figure}

We will proceed by using the Stokes variables defined as
\begin{equation}  \label{eq:StokesVariables}
\begin{split}
& S_0 = |\Omega_1|^2 + |\Omega_2|^2,\quad
S_3 = |\Omega_1|^2 - |\Omega_2|^2,\\
& S_1 + i S_2 = 2\bOmega_1 \Omega_2. 
\end{split}
\end{equation}
The following result will prove central,
\begin{equation} \label{eq:S1S2_timeDerivative}
\frac{d}{dt} (S_1+i S_2) = -\frac{i\,J}{4}\,\Mminus\, (S_1+2 + i S_2)^2.
\end{equation}
If we define, $w=S_1+2 + i S_2$, then Eq.~\eqref{eq:S1S2_timeDerivative} is
\begin{equation} \label{eq:dwdt}
\frac{dw}{dt} = i\gamma\,w^2
\end{equation}
where $\gamma$ is implicitly defined.
This equation gives invariant circles $w\bar{w} - R (w+\bar{w}) = 0$, or
\begin{equation} \label{eq:S1S2_circle}
    (\St_1+2-R)^2 + \St_2^2 = R^2,
\end{equation}
where $R\in\mathbb{R}$ is an arbitrary constant.
Fig.~\ref{fig:S1S2_circle} shows examples of these circles.
Solving for $R$, we find that a conserved quantity is explicitly written as
\begin{equation} \label{eq:circle_offOrigin}
\frac{(\St_1+2)^2 + \St_2^2}{\St_1 + 2}  = 2R.
\end{equation}

Eq.~\eqref{eq:circle_offOrigin} reproduces the result in Eq.~\eqref{eq:conserved1_exchange}.
In order to see this, we write \eqref{eq:conserved1_exchange} in terms of the Stokes variables,
\begin{equation} \label{eq:conserved1_Stokes}
I_1 = J \frac{(\St_1+2)^2 + \St_2^2}{\St_1^2+\St_2^2-4}.
\end{equation}
Both Eqs.~\eqref{eq:circle_offOrigin} and \eqref{eq:conserved1_Stokes} are quadratic in $\St_1, \St_2$, and they give the same family of circles.

We conclude this subsection with some remarks.
The point $S_1 + i S_2 = -2$ has a special role in this formulation, as it gives a fixed point of Eq.~\eqref{eq:S1S2_timeDerivative}.
It corresponds to $\bOmega_1 \Omega_2 = -1 \Leftrightarrow \magn_1 = -\magn_2$.

If we consider anisotropy $\anisotropy\neq 0$, the integral \eqref{eq:conserved1} does not represent simply a curve on the $(\St_1, \St_2)$ plane but a more complicated surface in the space $(\St_1, \St_2, \St_3)$.

A comparison of the results of the present section can be made with the more extensively studied system of two coupled nonlinear Schr\"odinger equations with cubic nonlinearity.
The system has two conserved quantities that were reported
in \cite{2010_PRA_RamezaniKottos}, and more extensively explained in
Refs.~\cite{2013_PRA_BarashenkovJackson,2013_PRA_PicktonSusanto,2013_JPhys_KevrekidisPelinovsky}.
A calculation similar to that in Eq.~\eqref{eq:S1S2_timeDerivative} gives a form $dw/dt = i\gamma'\,w$ with $\gamma'$ some quantity independent of $w$; cf. Eq.~\eqref{eq:dwdt}.
The solutions are a family of circles with the center at the origin of the plane $(\St_1, \St_2)$.

\subsection{Exchange model, second integral}
\label{sec:secondIntegral}

\begin{figure}[t]
    \centering
    \includegraphics[width=7cm]{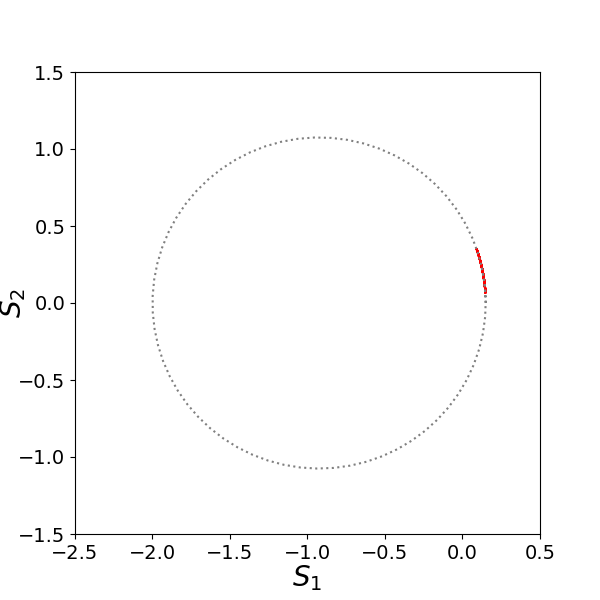}
    \caption{We simulate Eqs.~\eqref{eq:twoSpins_Omega} with parameters $J=1$ and $\storque=0.8$ (we set $\anisotropy=0, \hext=0$ and $\alpha=0$).
    The grey dotted line shows the circle \eqref{eq:circle_offOrigin} and the red arc shows the trajectory on the $(\St_1, \St_2)$ plane.
    All Stokes variables present oscillating motion in time (not shown).
    }
    \label{fig:S1S2_circle_k00b08}
\end{figure}

We write Eq.~\eqref{eq:S1S2_circle} in the parametric form
\begin{equation} \label{eq:S1S2_circle_parametric}
S_1 + i S_2 = R - 2 + R e^{i\phi},\qquad 0 \leq \phi < 2\pi,
\end{equation}
whose geometric meaning is shown in Fig.~\ref{fig:S1S2_circle}.
We substitute Eq.~\eqref{eq:S1S2_circle_parametric} in Eq.~\eqref{eq:S1S2_timeDerivative} and obtain
\begin{equation} \label{eq:phi_timeDerivative}
    \frac{d}{dt} \left[\tan\left( \phi/2 \right)\right] = \frac{R}{2}J\,\Mminus.
\end{equation}
Eq.~\eqref{eq:phi_timeDerivative} gives the rate at which we move on a circle defined in Eq.~\eqref{eq:S1S2_circle}.
An example of the trajectory during the motion on the $(\St_1, \St_2)$ plane is shown in Fig.~\ref{fig:S1S2_circle_k00b08}.

We can now combine Eq.~\eqref{eq:phi_timeDerivative} with Eq.~\eqref{eq:Mz_timeDerivative} and obtain a second conserved quantity
\begin{equation} \label{eq:conserved2}
I_2 = \tan\left( \phi/2 \right) + \frac{R}{2}\frac{J}{\storque}\,\ln|M_z|.
\end{equation}
In Fig.~\ref{fig:S1S2_circle}, we see that
\[
\tan(\phi/2) = \frac{\sin\phi}{1+\cos\phi} = \frac{S_2}{S_1+2}.
\]
and the conserved quantity \eqref{eq:conserved2} is written in terms of the Stokes variables as
\begin{equation} \label{eq:conserved2_stokes}
I_2 = \frac{S_2}{S_1+2} + \frac{R}{2}\frac{J}{\storque}\, \ln \left| 2 \frac{4-(S_1^2+S_2^2)}{(S_0+2)^2-S_3^2} \right|
\end{equation}
where we have used
\[
M_z = 2 \frac{4-S_0^2+S_3^2}{(S_0+2)^2-S_3^2} = 2 \frac{4-(S_1^2+S_2^2)}{(S_0+2)^2-S_3^2}.
\]

The integral \eqref{eq:conserved2_stokes} could also be obtained by a combination of Eqs.~\eqref{eq:Mz_timeDerivative} and \eqref{eq:S1S2_timeDerivative}.
However, the method used in this subsection is a more transparent one as it is based on geometric arguments.

Finally, we note that if the system of equations could be produced by a Hamiltonian, the existence of the two integrals of motion would imply its complete integrability \cite{2015_JPA_BarashenkovPelinovsky}.

\section{Concluding remarks}
\label{sec:conclusions}

A system of interacting spins that are under spin transfer torque can be described by a complex function, or a non-Hermitian Hamiltonian.
We give the formalism for obtaining the complex function that makes manifest the $\mathcal{PT}$ symmetry when this is present.
We have studied the nonlinear dynamics dynamics and the exceptional point for a system of two exchange-coupled spins in a case of $\mathcal{PT}$ symmetry.
This introduces a paradigm for the dynamics of non-Hermitian systems defined on the sphere. 

We have identified the regime for periodic precessional motion of the two spins, and its stability.
The periodic motion corresponds to a locking of the complete system in sustained magnetization oscillations. 
The synchronization of the oscillations of the two spins, due to $\mathcal{PT}$ symmetry, could lead to an answer to the long standing problem of the synchronization of spin-transfer torque nano-oscillators (STNO) \cite{2017_NatPhys_AwadAkerman,2017_NatComms_LebrunCros}.
The rich spin dynamics for the spin system can be further used to study non-magnetic systems which are described by effective spin variables (e.g., as in polariton condensates \cite{2020_NPJ_GhoshLiew}).

\section*{Acknowledgements}
The author is grateful to Kostas Makris for discussions and insightful remarks.
This work was supported by the project “ThunderSKY” funded by the Hellenic Foundation for Research and Innovation and the General Secretariat for Research and Innovation, under Grant No. 871.

\bigskip


\end{document}